\documentclass[journal]{IEEEtran}
\usepackage{amsmath,amssymb,amsfonts}
\usepackage{graphicx}
\usepackage{hyperref}
\usepackage{cleveref}
\usepackage{amssymb}
\usepackage{subcaption}
\usepackage{textcomp}
\usepackage{tabularx}
\usepackage{array}
\usepackage{makecell}
\usepackage{xcolor}
\usepackage{stackengine}
\usepackage{lipsum}
\usepackage[linesnumbered,ruled,vlined]{algorithm}
\usepackage{listings}
\usepackage[noend]{algpseudocode}
\newcommand*{\rom}[1]{\expandafter\@slowromancap\romannumeral #1@}
\makeatother

\usepackage[compact]{titlesec}
\titlespacing{\section}{1.0pt}{*1.0}{*0}
\titlespacing{\subsection}{1.1pt}{*1.1}{*0}
\titlespacing{\subsubsection}{0.3pt}{*0}{*0}

\usepackage{subcaption}
\usepackage{caption}

\definecolor{GreenForest}{rgb}{0.09, 0.45, 0.27}
\usepackage{soul}
\usepackage{balance}

\usepackage{cite}
\usepackage{amssymb}
\usepackage{acro}

\DeclareAcronym{GNSS}{
  short = GNSS,
long  = global navigation satellite system,
  tag = abbrev
}

\DeclareAcronym{UE}{
  short = UE,
long  = user equipment,
  tag = abbrev
}
\DeclareAcronym{CSI}{
  short = CSI,
long  = channel state information,
  tag = abbrev
}

\DeclareAcronym{ISAC}{
  short = ISAC,
  long  = integrated sensing and communication,
  tag = abbrev
}
\DeclareAcronym{CFO}{
  short = CFO,
  long  = carrier frequency offset,
  tag = abbrev
}

\DeclareAcronym{NIC}{
  short = NIC,
  long  = network interface card,
  tag = abbrev
}

\DeclareAcronym{TDOA}{
  short = TDOA,
  long  = time difference of arrival,
  tag = abbrev
}
\DeclareAcronym{FMCW}{
  short = FMCW,
  long  = Frequency Modulated Continuous Wave,
  tag = abbrev
}
\DeclareAcronym{LTE-V}{
  short = LTE-V,
  long  = long term evolution-vehicle,
  tag = abbrev
}

\DeclareAcronym{DPD}{
  short = DPD,
  long  = direct position determination,
  tag = abbrev
}

\DeclareAcronym{TOA}{
  short = TOA,
  long  = time of arrival,
  tag = abbrev
}
\DeclareAcronym{NTN}{
  short = NTN,
  long  = non terrestrial networks,
  tag = abbrev
}

\DeclareAcronym{DFO}{
  short = DFO,
  long  = Doppler frequency offset,
  tag = abbrev
}
\DeclareAcronym{V2X}{
  short = V2X,
  long  = vehicle to every thing,
  tag = abbrev
}

\DeclareAcronym{PPM}{
  short = PPM,
  long  = parts per million,
  tag = abbrev
}

\DeclareAcronym{ML}{
  short = ML,
  long  = maximum likelihood,
  tag = abbrev
}
\DeclareAcronym{SNR}{
  short = SNR,
  long  = signal to noise ratio,
  tag = abbrev
}

\DeclareAcronym{SDR}{
  short = SDR,
  long  = software defined radio,
  tag = abbrev
}

\DeclareAcronym{V2V}{
  short = V2V,
  long  = vehicle to vehicle,
  tag = abbrev
}
\DeclareAcronym{CP}{
  short = CP,
  long  = cyclic prefix  ,
  tag = abbrev
}

\DeclareAcronym{BS}{
  short = BS,
  long  = base station  ,
  tag = abbrev
}

\DeclareAcronym{gNB}{
  short = gNB,
  long  = next-generation NodeB,
  tag = abbrev
}

\DeclareAcronym{TRP}{
  short = TRP,
  long  = transmission-reception point,
  tag = abbrev
}

\DeclareAcronym{PDSCH}{
  short = PDSCH,
  long  = physical downlink shared channel,
  tag = abbrev
}

\DeclareAcronym{PUSCH}{
  short = PUSCH,
  long  = physical uplink shared channel,
  tag = abbrev
}

\DeclareAcronym{SRS}{
  short = SRS,
  long  = sounding reference signal,
  tag = abbrev
}

\DeclareAcronym{RE}{
  short = RE,
  long  = resource element,
  tag = abbrev
}

\DeclareAcronym{ToF}{
  short = ToF,
  long  = time of flight  ,
  tag = abbrev
}

\DeclareAcronym{AoA}{
  short = AoA,
  long  = Angle of Arrival,
  tag = abbrev
}

\DeclareAcronym{AoD}{
  short = AoD,
  long  = Angle of Departure,
  tag = abbrev
}

\DeclareAcronym{ISI}{
  short = ISI,
  long  = inter-symbol interference ,
  tag = abbrev
}

\DeclareAcronym{IFFT}{
  short = IFFT,
  long  = inverse fast Fourier transform ,
  tag = abbrev
}

\DeclareAcronym{RMSE}{
  short = RMSE,
  long  = root mean square error ,
  tag = abbrev
}

\DeclareAcronym{HW}{
  short = HW,
  long  = hardware,
  tag = abbrev
}

\DeclareAcronym{IIOT}{
  short = IIOT,
  long  = industrial internet of things,
  tag = abbrev
}
\DeclareAcronym{IoT}{
  short = IoT,
  long  = internet of things,
  tag = abbrev
}

\DeclareAcronym{OFDM}{
  short = OFDM,
  long  = orthogonal frequency division multiplexing,
  tag = abbrev
}
\DeclareAcronym{URLLC}{
  short = URLLC,
  long  = ultra reliable low latency communication,
  tag = abbrev
}

\DeclareAcronym{3GPP}{
  short = 3GPP,
  long  = 3rd generation partnership project ,
  tag = abbrev
}

\DeclareAcronym{NLOS}{
  short = NLOS,
  long  = non line of sight,
  tag = abbrev
}
\DeclareAcronym{LOS}{
  short = LOS,
  long  = line of sight,
  tag = abbrev
}

\DeclareAcronym{TDD}{
  short = TDD,
  long  = time division duplexing,
  tag = abbrev
}

\DeclareAcronym{PDP}{
  short = PDP,
  long  = power delay profile,
  tag = abbrev
}
\DeclareAcronym{ICI}{
  short = ICI,
  long  = inter-carrier-interference,
  tag = abbrev
}
\DeclareAcronym{RSU}{
  short = RSU,
  long  = road side unit,
  tag = abbrev
}

\DeclareAcronym{PRS}{
  short = PRS,
  long  = positioning reference signal,
  tag = abbrev
}
\DeclareAcronym{CPR}{
  short = CPR,
  long  = common phase rotation,
  tag  = abbrev
}

\DeclareAcronym{OTA}{
  short = OTA,
  long  = over the air,
  tag  = abbrev
}

\DeclareAcronym{CP-OFDM}{
  short = CP-OFDM,
  long  = cyclic prefix orthogonal frequency division multiplexing,
  tag  = abbrev
}
\DeclareAcronym{FFT}{
  short = FFT,
  long  = fast Fourier transform,
  tag  = abbrev
}
\DeclareAcronym{PN}{
  short = PN,
  long  = phase noise,
  tag  = abbrev
}

\DeclareAcronym{CPE}{
  short = CPE,
  long  = common phase error,
  tag  = abbrev
}

\DeclareAcronym{LO}{
  short = LO,
  long  = local oscillator,
  tag = abbrev
}

\DeclareAcronym{AWGN}{
  short = AWGN,
  long  = additive white Gaussian noise,
  tag = abbrev
}

\DeclareAcronym{ULA}{
  short = ULA,
  long  = uniform linear array,
  tag = abbrev
}

\DeclareAcronym{RTT}{
  short = RTT,
  long  = round-trip time,
  tag = abbrev
}

\DeclareAcronym{TDoA}{
  short = TDoA,
  long  = time difference of arrival,
  tag = abbrev
}

\DeclareAcronym{UL}{
  short = UL,
  long  = uplink,
  tag = abbrev
}

\DeclareAcronym{DL}{
  short = DL,
  long  = downlink,
  tag = abbrev
}

\DeclareAcronym{TDL}{
  short = TDL,
  long  = tap delay line,
  tag = abbrev
}

\DeclareAcronym{PTRS}{
  short = PTRS,
  long  = phase tracking reference signal,
  tag = abbrev
}

\DeclareAcronym{CRLB}{
  short = CRLB,
  long  = Cram\'{e}r--Rao lower bound,
  tag = abbrev
}

\DeclareAcronym{ICFO}{
  short = ICFO,
  long  = integer carrier frequency offset,
  tag = abbrev
}

\DeclareAcronym{FCFO}{
  short = FCFO,
  long  = fractional carrier frequency offset,
  tag = abbrev
}

\DeclareAcronym{SL}{
  short = SL,
  long  = sidelink,
  tag = abbrev
}

\DeclareAcronym{RF}{
  short = RF,
  long  = radio frequency,
  tag = abbrev
}

\DeclareAcronym{Uu}{
  short = Uu,
  long  = user equipment - gNB radio interface,
  tag   = abbrev
}

\DeclareAcronym{CDF}{
  short = CDF,
  long  = cumulative distribution function,
  tag = abbrev
}

\DeclareAcronym{LS}{
  short = LS,
  long  = least squares,
  tag = abbrev
}

\DeclareAcronym{FEC}{
  short = FEC,
  long  = forward error correction,
  tag = abbrev
}

\DeclareAcronym{CRB}{
  short = CRB,
  long  = Cram\'{e}r--Rao bound,
  tag = abbrev
}

\DeclareAcronym{NR}{
  short = NR,
  long  = New Radio,
  tag = abbrev
}

\begin{document}
    \title{Bidirectional CFO Separation for Radial Velocity Estimation in 5G NR TDD V2X Links}
  \author{ Mohamed Elamine Benattia, and H\"{u}seyin Arslan,~\IEEEmembership{Fellow,~IEEE}
\thanks{Mohamed Elamine and Hüseyin Arslan are with the Department of Electrical and Electronics Engineering, Istanbul Medipol University, Istanbul, 34810, Turkey (e-mail: Mohamed.benattia@std.medipol.edu.tr; huseyinarslan@medipol.edu.tr).}}
\maketitle
\begin{abstract}
High-accuracy velocity estimation over orthogonal frequency division multiplexing (OFDM) links is challenging because Doppler shifts and local oscillator (LO) drift contribute the same way to the observed carrier frequency offset (CFO). We propose a bidirectional CFO separation method for a frequency range-1 (FR1) time division duplex (TDD) Uu link between a roadside gNB/TRP and a vehicle user equipment (UE). Downlink (DL) and uplink (UL) CFO estimates are obtained from the physical downlink and uplink shared channel phase tracking reference signals (PT-RS), respectively, their sum isolates the radial Doppler and allow for the extraction of the radial velocity, while their difference recovers the relative LO offset. The estimates are obtained in closed form, without any iterative processing, using only a compact scalar CFO report fed back on the UL, the method can be co-scheduled with an new radio (NR) round trip time (RTT) positioning event to enable simultaneous ranging measurements along velocity estimates. We derive an idealized Cram\'er-Rao lower bound (CRLB), show that a linear phase slope estimator attains it at high signal to noise ratio, evaluate sensitivity to angle of arrival uncertainty and multipath, and discuss RF chain mismatch. The resulting link level implementation is aligned with NR PDSCH/PUSCH PT-RS procedures and is suitable for network assisted vehicle to everything (V2X) integrated sensing and communication (ISAC).
\end{abstract}
\begin{IEEEkeywords}
Carrier frequency offset, Doppler estimation, oscillator drift, radial velocity, integrated sensing and communication, OFDM, V2X, 5G NR, Uu, PT-RS, RTT.
\end{IEEEkeywords}
\IEEEpeerreviewmaketitle

\section{Introduction}

\Ac{OFDM} links can estimate radial velocity from Doppler shifts over standard communication links~\cite{niu2022rethinking}. Before demodulation, receivers compensate the \ac{CFO}, which combines Doppler with hardware induced \ac{LO} drift from both \ac{RF} chains. Conventional estimators treat this composite \ac{CFO} as one parameter, which is sufficient for communication but causes large sensing errors at high carrier frequencies or extreme speeds because Doppler cannot be separated from hardware bias.

\begin{figure}[h!]
    \centering
    \includegraphics[width=0.7\linewidth]{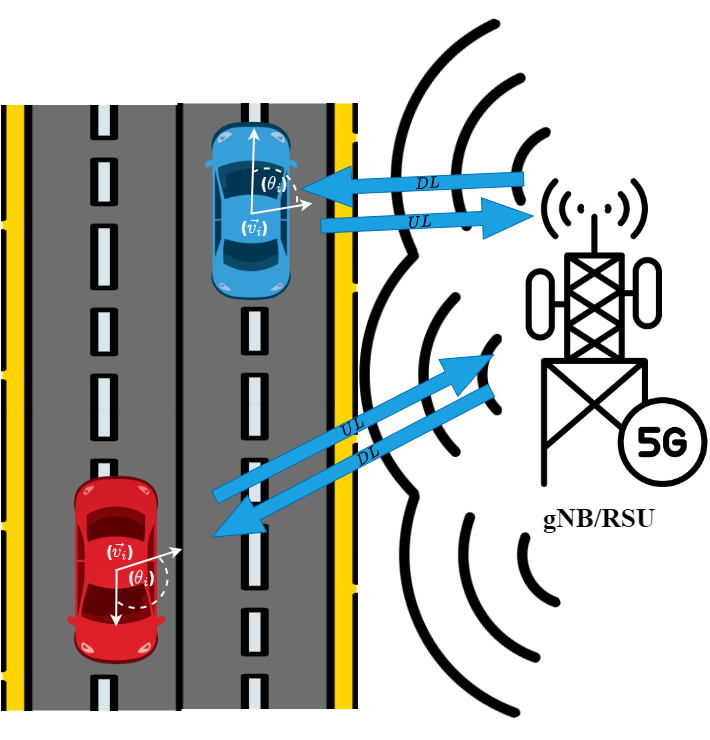}
    \caption{FR1 \ac{TDD} \ac{Uu} \ac{V2X} scenario between a roadside \ac{gNB}/\ac{TRP} (\ac{RSU}) and vehicle \acp{UE} traveling at speed $v_i$ with relative horizontal angle $\theta_i$ between the velocity vector and the \ac{LOS}.}
    \label{fig:5G_SL}
\end{figure}

\begin{figure}
    \centering
    \includegraphics[width=0.9\linewidth]{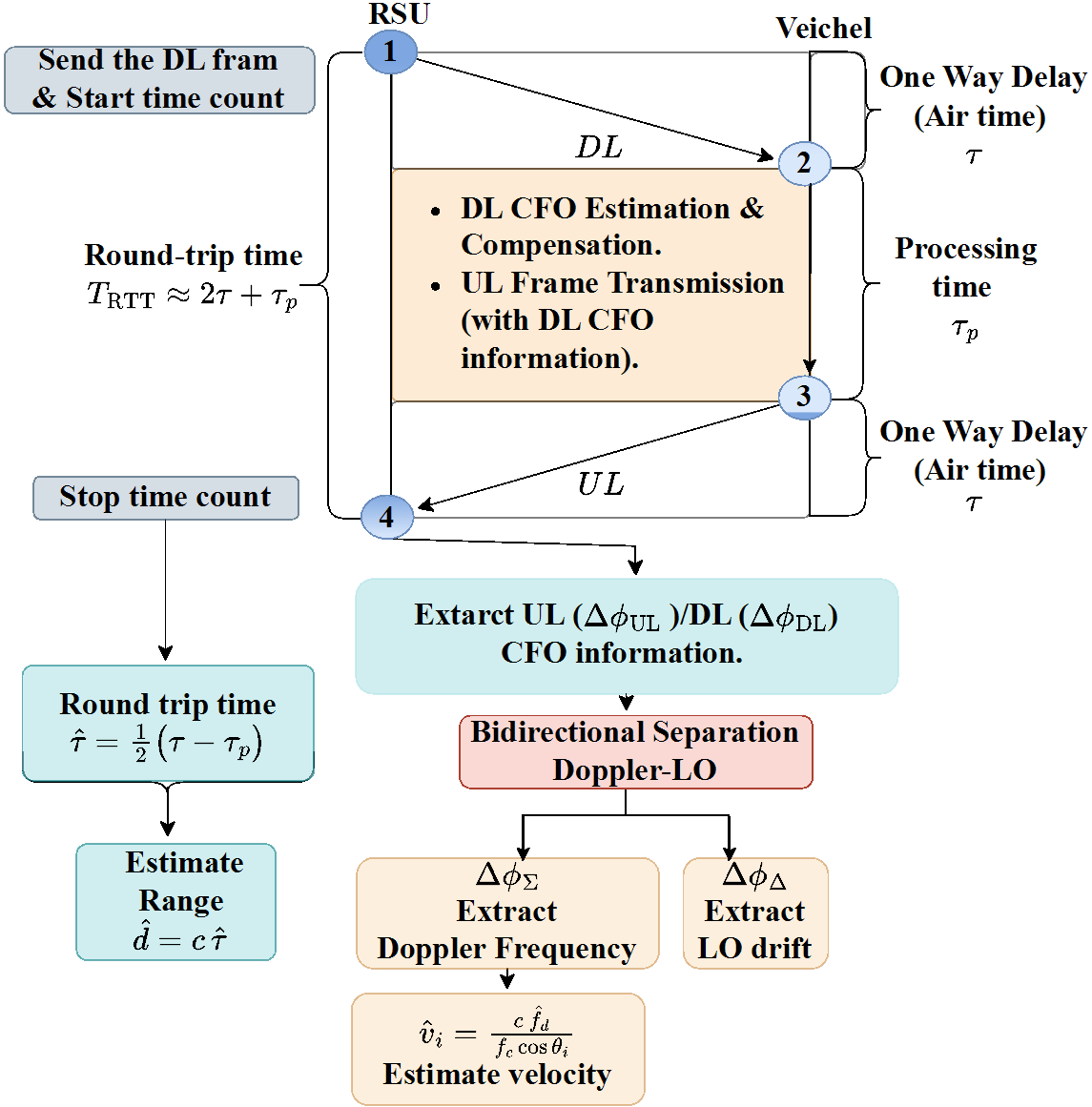}
    \caption{Flow of the proposed scheme to get the range and velocity information via bidirectional \ac{CFO} estimation and the \ac{RTT}-based ranging procedure.}
    \label{fig:Flow}
\end{figure}

Existing asynchronous \ac{ISAC} methods resolve this ambiguity only under additional constraints. Monostatic \ac{OFDM} radars use a shared \ac{LO} and iterative joint estimation to remove bias and phase noise~\cite{keskin2023monostatic}. Bistatic and distributed systems use static reference paths or dominant static reflections to cancel the offset common to all paths~\cite{pegoraro2024jump,ventura2024bistatic}, while clock asynchronous methods require multiple frames or receivers to preserve Doppler coherence~\cite{wu2024sensing}. Other methods jointly estimate frequency offset and Doppler using an orthogonal angle domain subspace~\cite{zeng2018joint}, use \ac{CFO} for \ac{TDoA} localization~\cite{hannotier2024cfo}, or jointly estimate Doppler spread and \ac{CFO} in flat fading channels using \ac{ML}~\cite{Bellili2017TCOM}. These approaches require spatially separable paths, synchronized anchor networks, or estimate Doppler spread rather than the signed radial Doppler needed for velocity. For 5G \ac{NR} \ac{V2X}, prior work studies sub-6-GHz \ac{SL} positioning~\cite{Analysis_of_V2X_Sidelink_6Ghz} and \ac{OFDM} localization under hardware impairments with dedicated bounds~\cite{HWI-5G-6G}, but treats oscillator offset as a calibrated nuisance rather than an identifiable parameter. In time based ranging, clock drift is the main impairment, where double sided \ac{RTT} cancels it by transmitting twice, at the cost of latency, as in classical double sided two way ranging that removes \ac{PPM} clock drift between unsynchronized devices~\cite{gao2026sidelink,neirynck2016}. However, these methods discard the clock term after protecting the delay estimate. In this work, we retain both directions of a single exchange to jointly identify Doppler and oscillator offset.\\
We consider a network connected roadside \ac{gNB}/\ac{TRP}, we use the scenario of a \ac{RSU}, and a vehicle \ac{UE} over an FR1 \ac{TDD} \ac{Uu} link. We apply a method called bidirectional cancellation that is long used in two way noncoherent satellite and deep space Doppler tracking to a single \ac{TDD} exchange, the sign reversal of the \ac{LO} term between the \ac{DL} and \ac{UL} \ac{CFO} algebraically separates the Doppler shift from the \ac{LO} induced bias (Figs.~\ref{fig:5G_SL},~\ref{fig:Flow}). To the best of our knowledge, this bidirectional \ac{CFO} decomposition has not previously been applied to \ac{PDSCH}/\ac{PUSCH} \ac{PTRS} observations for radial velocity estimation in an \ac{NR} \ac{TDD} \ac{V2X} compatible link, it requires no external reference signals, no iterative processing, and only a compact scalar \ac{CFO} report fed back on the \ac{UL}. We derive a closed form \ac{CRLB} for the joint Doppler/\ac{LO} drift estimation, show that a linear phase slope estimator attains it, and evaluate it over \ac{V2X} channels against multipath, \ac{AoA} error, and \ac{RF} chain mismatch.
\section{System Model and Signal Formulation}
\subsection{OFDM Signal Model}
An \ac{OFDM} transmitter maps frequency domain symbols $X_{k,m}$ onto subcarriers $k\in\{0,\dots,N-1\}$ for each symbol $m\in\{0,\dots,M-1\}$ in an $M\times N$ frame.
The total bandwidth is $B$ with subcarrier spacing $\Delta f=B/N$, useful symbol duration $T=1/\Delta f$, \ac{CP} length $T_{\mathrm{cp}}$, and total symbol time $T_s=T+T_{\mathrm{cp}}$.
The baseband transmit signal is
\begin{equation}
\label{eq:tx}
x(t)=\frac{1}{\sqrt{N}}\sum_{m=0}^{M-1}\sum_{k=0}^{N-1}
X_{k,m}\,e^{j2\pi k\Delta f (t-mT_s)}\,
\mathrm{rect}\!\Big(\frac{t-mT_s}{T_s}\Big),
\end{equation}
where $\mathrm{rect}(t)=1$ for $|t|\le \tfrac12$ and $0$ otherwise. Pilot placements on $(k,m)$ are assumed known.
\subsection{Reference Signals}
The \ac{CFO} estimator uses \ac{PTRS} associated with \ac{PDSCH} and \ac{PUSCH} transmissions~\cite{3GPP_TS_38211}. \ac{PTRS} is designed to track phase noise and \ac{CPE} and provides known phase observations at configured time and frequency densities, suitable for the per symbol phase increment observations of Section~\ref{sec:proposed}, we use an inspired comb pilot pattern that spans the entire time domaine signal instead of strictly following the standards version. The accompanying \ac{NR} positioning procedure may use \ac{DL} \ac{PRS} and positioning \ac{SRS} for \ac{RTT} measurements~\cite{3GPP-Release-I8}, these positioning signals are separate from the \ac{PTRS} observations used by the proposed \ac{CFO} estimator.
\subsection{Doppler and Oscillator Model}
\label{sub_sec:channel}
\subsubsection{Wireless Channel and Mobility Effect}
A vehicle $i$ at a distance $d_i$ from the \ac{RSU} has one way delay $\tau_i=d_i/c$, where $c$ is the speed of light. The vehicle moves with speed $v_i$ at an angle $\theta_i$ to the \ac{RSU} line of sight, giving radial velocity $v_{r,i}=v_i\cos\theta_i$. The \ac{AoA} can be estimated via array processing and is assumed known. Section~\ref{sec:AoA} analyzes the impact of \ac{AoA} errors on the full speed estimate. The signal at carrier $f_c$ then experiences the Doppler shift
\begin{equation}
\label{eq:velo}
f_{d,i}=\frac{v_{r,i}\,f_c}{c}=\frac{v_i f_c \cos\theta_i}{c}.
\end{equation}
We adopt a synthetic Rician tapped delay channel with $L$ paths and Rician $K$ factor, parameterised with in the \ac{3GPP} framework~\cite{3GPP_TR_38.901_v15.0.0}\footnote{A dominant \ac{LOS} term plus equal power diffuse taps, rather than a full clustered \cite{3GPP_TR_38.901_v15.0.0} realisation}, described as
\begin{equation}
\label{eq:channel}
h(t,\tau)=\alpha_0\,e^{j2\pi f_{d,0} t}\delta(\tau-\tau_0)+\sum_{\ell=1}^{L-1}\alpha_\ell\,e^{j2\pi (f_{d,\ell}) t}\,\delta(\tau-\tau_\ell),
\end{equation}
where the \ac{LOS} path ($\ell=0$, gain $\alpha_0$, delay $\tau_0$) carries the dominant Doppler $f_{d,0}=f_{d,i}$, and each \ac{NLOS} tap carries its Doppler relative to the \ac{LOS}. The \ac{LOS} Doppler is absorbed into the aggregate offset $\varepsilon$ applied in Section~\ref{sec:reception}, while the residual \ac{NLOS} terms perturb the pilot phase. Section~\ref{sec:multipath} evaluates the impact of this multipath on the proposed estimator.
\subsubsection{Oscillator Model}
All practical oscillators exhibit a quasi static residual \ac{CFO} together with stochastic phase noise due to \ac{LO} drift~\cite{leeson2016oscillator}. We keep the distinction of the residual \ac{CFO} as the deterministic frequency term, while the \ac{PN} is modeled as a Wiener process~\cite{zou2007phase},
\begin{equation}
\Phi(t+\Delta t)=\Phi(t)+\eta(t),\qquad 
\eta(t)\sim\mathcal{N}(0,\sigma_\Phi^2\,\Delta t),
\end{equation}
where $\eta(t)$ is the stochastic innovation term that follows a normal distribution. Let $\Phi_{\rm tx}(t)$ and $\Phi_{\rm rx}(t)$ denote the transmitter and receiver \ac{PN} processes, with their difference modeled as $\Phi(t)\triangleq\Phi_{\rm tx}(t)-\Phi_{\rm rx}(t)$ \cite{david2015tracking}. Let the roadside \ac{gNB}/\ac{TRP} ($B$) and vehicle \ac{UE} ($V$) \ac{LO} have residual frequency errors $\epsilon_{B,i}$ and $\epsilon_{V,i}$, respectively. The effective estimated \ac{CFO} on the \ac{DL} and \ac{UL} are described as
\begin{equation}
\label{eq:Diffrance_LO}
\epsilon_{BV,i}=\epsilon_{B,i}-\epsilon_{V,i},
\qquad
\epsilon_{VB,i}=-\epsilon_{BV,i},
\end{equation}
which are assumed to remain approximately constant over a single \ac{RTT} window. This window spans the observation frame of a few milliseconds for the numerology of Table~\ref{tab:Simulation_Parameters}, over which a $1$\,ppm class oscillator drifts negligibly, so the residual \ac{CFO} is quasi static. Adding the Doppler shift $f_{d,i}$ yields the total frequency offsets
\begin{equation}
\varepsilon_{\mathrm{DL},i}=f_{d,i}+\epsilon_{BV,i},
\qquad
\varepsilon_{\mathrm{UL},i}=f_{d,i}-\epsilon_{BV,i}.
\end{equation}

\section{Downlink and Uplink Reception}
\label{sec:reception}
The received signal under Doppler and \ac{LO} drift effects is modeled as
\begin{equation}
r(t)= e^{j(2\pi \varepsilon t + \Phi(t))} \bigl(x(t) * h(t)\bigr)+ w(t),
\end{equation}
where $w(t)$ is complex \ac{AWGN} with mean ($\mu=0$) and variance $\sigma^2$. Here the aggregate offset $\varepsilon$ carries the \ac{LOS} Doppler and the \ac{LO} offset, consistent with the \ac{LOS}/\ac{NLOS} split of \eqref{eq:channel}.
\subsection{Downlink Baseband Signal}
For the \ac{LOS} path with delay $\tau$ and gain $\alpha$, the \ac{DL} \ac{PDSCH} baseband signal at the $i$-th vehicle is
\begin{equation}
\label{eq:r_dl}
\begin{aligned}
r_{\mathrm{DL}}(t)
&= \alpha\,e^{j\phi(t)}\,e^{j2\pi \varepsilon_{\rm DL,i} t}
\sum_{m=0}^{M-1}\sum_{k=0}^{N-1} X_{k,m}\,e^{j2\pi k\Delta f (t-mT_s-\tau)}\\
&\quad\quad\times \mathrm{rect}\!\Big(\frac{t-\tau-mT_s}{T_s}\Big) + w(t).
\end{aligned}
\end{equation}
The vehicle estimates the \ac{DL} \ac{CFO} before frequency compensation, quantizes  $\hat\varepsilon_{\rm DL,i}$ and feeds it back to the \ac{gNB} as a compact scalar through an implementation specific \ac{UL} control message or data payload, the \ac{gNB} obtains the \ac{UL} \ac{CFO} from the \ac{PUSCH} \ac{PTRS}.

\subsection{Uplink Baseband Signal}
After a processing delay $\tau_p$, the vehicle transmits the \ac{UL} \ac{PUSCH} frame over the same reciprocal \ac{TDD} channel. The \ac{gNB} then receives $r_{\mathrm{UL}}(t)$, identical in form to \eqref{eq:r_dl} but with the offset $\varepsilon_{\rm UL,i}$ replacing $\varepsilon_{\rm DL,i}$, by \ac{TDD} reciprocity the \ac{UL} propagation delay equals the one way air time $\tau$, and is needed for \ac{RTT} timing of Section~\ref{sec:rtt}.
\section{Association With NR RTT Positioning}
\label{sec:rtt}
The bidirectional \ac{CFO} exchange can be co-scheduled with an \ac{NR} \ac{RTT} session. As intuition, a single link round trip obeys $T_{\mathrm{RTT}}\approx 2\tau+\tau_p$, with $\tau_p$ the processing turnaround, giving
\begin{equation}
\label{eq:RTT}
\hat\tau=\tfrac{1}{2}\big(T_{\mathrm{RTT}}-\tau_p\big),\qquad
\hat d=c\,\hat\tau.
\end{equation}
In standardized \ac{NR} \ac{RTT}, however, \ac{UE} and \ac{gNB}/\ac{TRP} Rx-Tx time difference measurements are obtained from \ac{DL} \ac{PRS} and positioning \ac{SRS} and combined by the positioning system~\cite{3GPP-Release-I8}. The \ac{CFO} decomposition is applied separately to each suitable \ac{TDD} link \ac{CFO} measurement, and the focuses of this work is on \ac{CFO} based radial velocity and \ac{LO}-offset estimation and does not evaluate range accuracy. 
\section{Frequency Domain Observation Model}
After sampling at $f_s$, timing synchronization, \ac{CP} removal and \ac{FFT}, the linear convolution becomes circular for $T_{\mathrm{cp}} > \max_{\ell}\tau_{\ell}$. Under normalized frequency offset $\nu=\varepsilon/\Delta f$ where \ac{ICI} is manageable on pilots, the pilot observations are
\begin{equation}
\label{eq:obs}
\begin{aligned}
R^{\mathrm{DL}}_{k,m} &\approx \alpha\,X_{k,m}\,e^{-j2\pi k\Delta f\,\tau}\,
e^{j\big(2\pi \varepsilon_{\rm DL,i}\, mT_s+\phi_m\big)} + n^{\mathrm{DL}}_{k,m},\\
R^{\mathrm{UL}}_{k,m} &\approx \alpha\,X_{k,m}\,e^{-j2\pi k\Delta f\,\tau}\,
e^{j\big(2\pi \varepsilon_{\rm UL,i}\, mT_s+\phi_m\big)} + n^{\mathrm{UL}}_{k,m},
\end{aligned}
\end{equation}
where $\phi_m$ is the per symbol \ac{CPE} induced by \ac{PN}, and $n^{(\cdot)}_{k,m}$ includes \ac{AWGN} and any residual \ac{ICI}.
Let the instantaneous pilot phases be $\phi^{\mathrm{DL}}_{k,m}\triangleq\angle R^{\mathrm{DL}}_{k,m}$ and $\phi^{\mathrm{UL}}_{k,m}\triangleq\angle R^{\mathrm{UL}}_{k,m}$. On the pilots they evolve linearly with the OFDM symbol index, the slope being the per-direction frequency offset
\begin{equation}
\label{eq:phase_linear}
\begin{aligned}
\phi^{\mathrm{DL}}_{k,m} &\approx 2\pi\,\varepsilon_{\mathrm{DL},i}\,mT_s + \psi^{\mathrm{DL}}_k + \eta^{\mathrm{DL}}_{k,m},\\
\phi^{\mathrm{UL}}_{k,m} &\approx 2\pi\,\varepsilon_{\mathrm{UL},i}\,mT_s + \psi^{\mathrm{UL}}_k + \eta^{\mathrm{UL}}_{k,m},
\end{aligned}
\end{equation}
where $\psi_k$ is a constant phase set by the propagation delay and the known initial pilot phase, we de-rotate them before the slope, and $\eta_{k,m}$ collects residual multipath from \ac{NLOS}, oscillator \ac{PN}, and \ac{AWGN}.

\section{The Proposed Bidirectional Doppler–LO Separation}
\label{sec:proposed}
For each link direction, the frequency offset is estimated as the slope of the linear pilot phase in \eqref{eq:phase_linear}.  A least squares fit of the pilot phase against the symbol index, averaged over the $N_p$ \ac{PTRS} subcarriers, gives the \ac{DL} offset
\begin{equation}
\label{eq:slope}
\hat\varepsilon_{\mathrm{DL},i}=\frac{1}{2\pi T_s\,N_p}\sum_{k=1}^{N_p}\frac{\sum_{m}(m-\bar m)\,\phi^{\mathrm{DL}}_{k,m}}{\sum_{m}(m-\bar m)^2}, ~~ \bar m=\frac{1}{M}\sum_{m}m,
\end{equation}
and identically the \ac{UL} offset $\hat\varepsilon_{\mathrm{UL},i}$ from $\phi^{\mathrm{UL}}_{k,m}$. In the considered configuration a phase observation is available in every \ac{OFDM} symbol, so the pilot times are $t_m=mT_s$, for time sparse \ac{PTRS} the actual pilot times replace $mT_s$ in \eqref{eq:slope}. Centring the symbol index about $\bar m$ removes the constant phase $\psi_k$, so \eqref{eq:slope} is the standard single tone frequency estimator, whose variance attains the Cramér–Rao bound at high \ac{SNR}~\cite{tretter1985,rife1974} as derived in Section~\ref{sec:CRLB}. Using the \ac{TDD} sign flip property $\epsilon_{VB,i}=-\epsilon_{BV,i}$ in \eqref{eq:Diffrance_LO}, the sum and difference of the two per‑direction estimates separate the Doppler shift from the \ac{LO} drift
\begin{equation}
\label{eq:estimators}
\hat \epsilon_{BV,i}=\frac{\hat\varepsilon_{\mathrm{DL},i}-\hat\varepsilon_{\mathrm{UL},i}}{2},\quad
\hat f_d=\frac{\hat\varepsilon_{\mathrm{DL},i}+\hat\varepsilon_{\mathrm{UL},i}}{2},\quad
\hat v_{r,i}=\frac{c\,\hat f_d}{f_c}.
\end{equation}
The directly identifiable quantity is the radial velocity $\hat v_{r,i}$, the full speed $\hat v_i=\hat v_{r,i}/\cos\hat\theta_i$ is recovered only when the movement direction \ac{AoA} is independently known. That recovery is singular as $|\cos\theta_i|\!\to\!0$ in broadside measurements, whereas $\hat v_{r,i}$ itself remains well defined. The method is thus best suited to mobility driven \ac{LOS} geometries with reliable \ac{AoA} accuracy for full speed recovery.

\section{Theoretical Analysis and Practical Considerations}
\label{sec:analysis}
This section derives the \ac{CRLB} for the joint estimation of Doppler $f_d$ and the residual \ac{LO} drift $\epsilon_{BV,i}$, and provides a sensitivity analysis on \ac{AoA} errors, the impact of multipath, and non-ideal channel reciprocity.

\subsection{Cramér-Rao Lower Bound}
\label{sec:CRLB}
Each per-direction offset in \eqref{eq:slope} is a linear phase slope observed on $N_p$ pilots over the $N_t$ \ac{PTRS} bearing symbols. For a single complex tone at per-\ac{RE} \ac{SNR} $\rho$ the slope fit frequency variance is bounded by the Rife-Boorstyn \ac{CRLB}~\cite{rife1974}, averaging over the $N_p$ pilots, the per-direction bound is
\begin{equation}
\label{eq:crlb_dir}
\mathrm{Var}(\hat\varepsilon_{\mathrm{DL},i})=\mathrm{Var}(\hat\varepsilon_{\mathrm{UL},i})\ge \frac{1}{2(2\pi)^2 T_s^2\,\rho\,N_p\,S},\;\; S=\sum_n (m_n-\bar m)^2,
\end{equation}
over the \ac{PTRS} symbol indices $m_n$ for equally spaced symbols $S=N_t(N_t^2-1)/12$. Here $\rho=10^{\mathrm{SNR}/10}$ is the post \ac{FFT} per-\ac{PTRS} \ac{RE} \ac{SNR}. Equation~\eqref{eq:crlb_dir} is an idealized \ac{AWGN} single tone bound. Phase noise is omitted from the present simulations, and is considered as \ac{CPE} common across subcarriers that adds a floor to \ac{CRLB} and calls for a hybrid \ac{CRLB}~\cite{barbieri2007crb}. Under the simulated \ac{AWGN}/\ac{LOS} conditions the \ac{PTRS} are known and the residual nuisance is negligible, so \eqref{eq:crlb_dir} is attained. Because the \ac{DL} and \ac{UL} noises are independent, the sum/difference in \eqref{eq:estimators} halves the variance,
\begin{equation}
\label{eq:crlb_var}
\mathrm{Var}(\hat f_d)=\mathrm{Var}(\hat\epsilon_{BV,i})\ge \frac{1}{4(2\pi)^2 T_s^2\,\rho\,N_p\,S} ,
\end{equation}
and the radial velocity bound follows by error propagation through \eqref{eq:estimators},
\begin{equation}
\label{eq:crlb_vel}
\mathrm{Var}(\hat v_{r,i}) \ge \left(\frac{c}{f_c}\right)^2 \mathrm{Var}(\hat f_d).
\end{equation}
Section~\ref{sec:simulation} verifies that the phase slope estimator attains this bound.

\subsection{AoA Error Propagation and Sensitivity} \label{sec:AoA} 
Velocity estimation \eqref{eq:estimators} depends on the \ac{AoA} through $\hat v_i=c\hat f_d/(f_c\cos\theta_i)$, so \ac{AoA} error $\tilde\theta_i=\hat\theta_i-\theta_i$ of variance $\sigma_\theta^2$ propagates to the velocity. Using $c\hat f_d/f_c=v_i\cos\theta_i$ from \eqref{eq:velo}, the first order sensitivity is
\begin{equation}
\frac{\partial \hat v_i}{\partial\theta_i}
=\frac{c\hat f_d}{f_c}\,\sec\theta_i\tan\theta_i=v_i\tan\theta_i,
\end{equation}
so an angular error of standard deviation $\sigma_\theta$ in radians induces a full speed error $\sigma_{v,\theta}\approx v_i\,|\tan\theta_i|\,\sigma_\theta$, which vanishes at $\theta_i=0$ and diverges at broadside ($|\cos\theta_i|\!\to\!0$). The radial velocity $\hat v_{r,i}=c\hat f_d/f_c$ is directly observable and \ac{AoA} free, we report it as the primary output and recover the full speed only with a reliable \ac{AoA}. For example, at $v=19.4$~m/s a $\sigma_\theta\!\approx\!2^\circ$ keeps the full speed error below $1$~m/s for $\theta\!\lesssim\!56^\circ$ and below $2$~m/s for $\theta\!\lesssim\!71^\circ$, within the $15$~m/s Release 19 Category 4 vehicle target at $95\%$ confidence~\cite{3gpp22137}. A quality study of the \ac{AoA} accuracy with the full speed reporting rule and its standardized are left to future work.

\subsection{Multipath Effects}
\label{sec:multipath}
The analysis above assumed a dominant \ac{LOS} component. \ac{NLOS} reflections perturb the pilot phase in \eqref{eq:phase_linear} through $\eta_{k,m}$. For high Ricean $K$, the \ac{LOS} dominates and the estimator stays close to the \ac{CRLB}, whereas as $K$ decreases the perturbation variance grows and performance departs from the bound (Section~\ref{sec:simulation}).

\subsection{Channel Reciprocity and RF Front End Mismatch}
The scheme relies on \ac{TDD} channel reciprocity, the physical multipath channel between the \ac{RSU} and the vehicle is assumed reciprocal within the coherence time, while the transmit and receive \ac{RF} chains have independent amplitude and phase responses~\cite{jiang2018reciprocity}. A constant \ac{RF} phase mismatch $\delta_{\mathrm{RF}}$ between the \ac{DL} and \ac{UL} chains only shifts the phase intercept $\psi_k$ in \eqref{eq:phase_linear} and is therefore removed by the centred slope fit in \eqref{eq:slope}, so it does not bias the frequency estimates. A bias arises only if the mismatch varies over the observation window as the effective residual inter chain frequency $\Delta f_{\mathrm{RF}}$, being direction specific it does not obey the \ac{CFO} sign flip and leaks $\Delta f_{\mathrm{RF}}$ into the difference statistic. Such variation is slow relative to the \ac{RTT} window and is further treated by reciprocity calibration using reference signals or \ac{OTA} feedback~\cite{jiang2018reciprocity}. We assume perfect \ac{RF} calibration in our simulations.

\section{Simulation}
\label{sec:simulation}
We evaluate the proposed scheme in a highway \ac{V2X} scenario over a Ricean channel (Section~\ref{sub_sec:channel}) using a standards aligned FR1 \ac{Uu} \ac{TDD} link-level configuration summarised in Table~\ref{tab:Simulation_Parameters}. The per direction offsets are obtained by the phase slope estimator \eqref{eq:slope} and combined by \eqref{eq:estimators}, and the Rife--Boorstyn CRLB \eqref{eq:crlb_var}--\eqref{eq:crlb_vel} is superimposed on every accuracy curve.

\begin{table}[h!]
\centering
\caption{Simulation parameters.}
\label{tab:Simulation_Parameters}
\begin{tabular}{|l|r|}
\hline
\textbf{Parameter} & \textbf{Value} \\
\hline
Carrier / $B_{oc}=N_{act}\Delta f$ & $3.5$\,GHz / $46.8$\,MHz \\ \hline
Subcarrier spacing ($\Delta f$) & $60$\,kHz ($\mu=2$) \\ \hline
FFT / active subcarriers ($N_{act}$) & $1024$ / $780$ \\ \hline
$f_s=N\Delta f$ / $T_{CP}$ / $T_{s}$ & $61.44$\,MHz / $\approx 1.19 \mu s$ / $17.86\,\mu$s \\ \hline
 M /PT-RS density & $140$ / $4$\ \\ \hline
Vehicle speed / radial Doppler & $19.4$\,m/s ($\approx\!226$\,Hz) \\ \hline
LO offset ($\epsilon_{BV}$) & $1$\,ppm ($\approx\!3.5$\,kHz) \\ \hline
Angle of arrival ($\theta$) & $0^\circ$ \\ \hline
Baseline chan / SNR & Rician $K\!=\!10$\,dB / $-10:2: 20$\,dB \\ \hline
Monte-Carlo (RMSE / CDF) & $1000$ / $5000$ \\ \hline
\end{tabular}
\end{table}

\subsection{Accuracy vs SNR}
\begin{figure}[!h]
    \centering
   \includegraphics[width=0.9\linewidth]{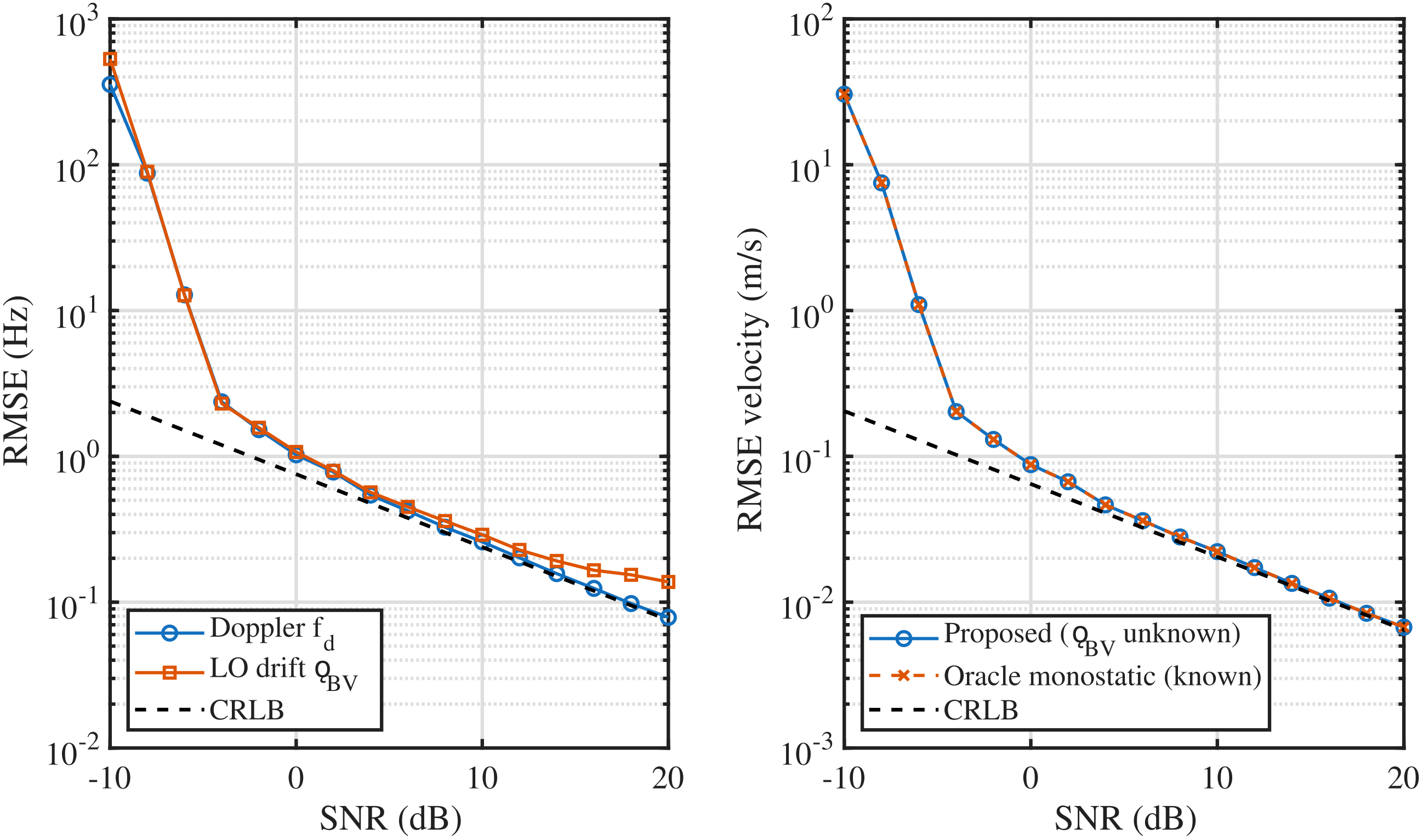}
    \caption{\ac{RMSE} vs \ac{SNR}: (left) Doppler and \ac{LO} drift of the proposed estimator, (right) velocity estimation of the proposed estimator VS a monostatic benchmark with known $\epsilon_{BV,i}$, both with the \ac{CRLB}.}
    \label{fig:oracle}
\end{figure}

Fig.~\ref{fig:oracle} plots the Doppler and \ac{LO} drift \ac{RMSE} vs \ac{SNR} (left) and the velocity \ac{RMSE} against a monostatic benchmark (right). The sum statistic (Doppler/velocity) attains the \ac{CRLB} for \ac{SNR}$\gtrsim8$~dB, with a sharp threshold roll off below $\approx0$~dB, while the difference statistic (\ac{LO} drift) tracks the bound but saturates at a small residual \ac{ICI} floor at high \ac{SNR} that does not effect the velocity estimation. The proposed estimator overlaps with the monostatic benchmark with known $\epsilon_{BV,i}$: jointly estimating Doppler with the unknown drift matches known \ac{LO} accuracy despite two independent oscillators, because the sum and difference statistics are uncorrelated under equal variance, and are independent in terms of \ac{DL}/\ac{UL} estimation.

\subsection{Identifiability, Robustness and Reliability}
\begin{figure}[!h]
    \centering
    \includegraphics[width=0.9\linewidth]{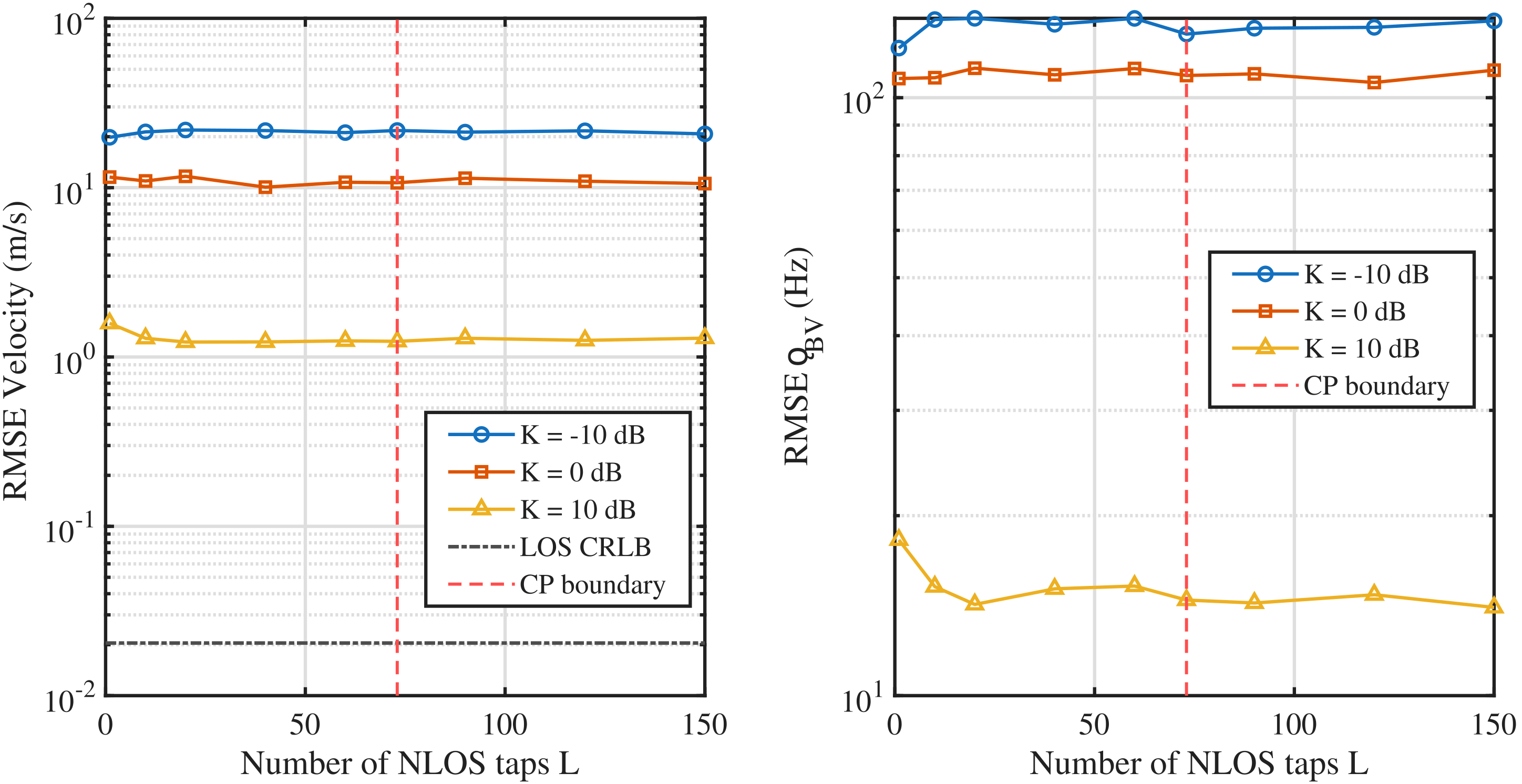}
    \caption{Velocity \ac{RMSE} (left) and \ac{LO} drift \ac{RMSE} (right) vs the number of \ac{NLOS} taps for $K=-10,0,10$~dB at \ac{SNR}$=10$~dB.}
    \label{fig:channel-performance}
\end{figure}

Fig.~\ref{fig:channel-performance} sweeps the number of \ac{NLOS} taps for three Ricean $K$ factors at \ac{SNR}$=10$~dB with exponential \ac{PDP}, each tap carries an independent Doppler, so the \ac{NLOS} perturbs the pilot phase by $\propto\!1/\sqrt{K}$. Both the velocity and \ac{LO} drift \ac{RMSE} are set almost entirely by $K$ and are nearly flat in the tap count, for velocity $\approx\!21,10,1.2$~m/s at $K=-10,0,10$~dB, because the exponential profile keeps the \ac{NLOS} power concentrated within the \ac{CP}, adding resolvable paths barely changes the perturbation, showing the system is more controled by the $K$ factor and best suited for \ac{LOS} dominant ($K\!\gtrsim\!0$~dB) geometries. 
The empirical error \acp{CDF}, a reliability oriented performance measure for \ac{ISAC} sensing~\cite{liu2024senscap}, give $99$th percentile errors at \ac{SNR}$=10$~dB of $0.056$~m/s for velocity and $0.73$~Hz for \ac{LO} drift. For context, \ac{3GPP} Release~19 specifies scenario dependent requirements for 5G wireless sensing, with $15$~m/s horizontal velocity accuracy at $95\%$ confidence for sensing vehicles and $99\%$ for public safety applications~\cite{3gpp22137}.

\section{Conclusion}
We presented a bidirectional \ac{CFO} decomposition for an \ac{NR} \ac{TDD} \ac{Uu} link between a roadside \ac{gNB}/\ac{TRP} and a vehicle \ac{UE}, combining \ac{DL} and \ac{UL} \ac{PTRS} phase observations to separate radial Doppler from the relative \ac{LO} offset without external references or iterative processing. A closed-form Rife-Boorstyn \ac{CRLB} was derived and shown to be attained by a linear phase slope estimator in the idealized \ac{AWGN} regime, with a radial velocity to $0.056$~m/s at the $99$th percentile at $10$~dB \ac{SNR}. The method may be co-scheduled with an \ac{NR} \ac{RTT} positioning session, while detailed ranging accuracy is outside the scope of this letter. Future work targets phase noise aware bounds, realistic vehicular channels, \ac{AoA} effects, \ac{SDR} and \ac{OTA} validation ,and full \ac{RTT} range evaluation.
\bibliographystyle{IEEEtran}
\bibliography{Bibliography}
\vfill
\end{document}